\documentclass{article}
\usepackage{tablefootnote}
\usepackage{tabularx}
\usepackage{makecell}
\usepackage{caption}
\usepackage{graphicx}
\usepackage{amsmath}
\usepackage{booktabs}
\usepackage{siunitx}
\usepackage{float}
\usepackage{amssymb}
\usepackage[margin=1in]{geometry}
\PassOptionsToPackage{hyphens}{url}\usepackage{hyperref}
\usepackage{xcolor}
\usepackage{tikz}
\usepackage[T1]{fontenc}

\title{Enhancing OHLC Data with Timing Features:\\A Machine Learning Evaluation}
\author{Ruslan Tepelyan \\ Bloomberg \\ New York, USA\\ \texttt{rtepelyan@bloomberg.net}}
\date{}

\begin{document}

\maketitle

\begin{abstract}
OHLC bar data is a widely used format for representing financial asset prices over time due to its balance of simplicity and informativeness. Bloomberg has recently introduced a new bar data product that includes additional timing information—specifically, the timestamps of the open, high, low, and close prices within each bar. In this paper, we investigate the impact of incorporating this timing data into machine learning models for predicting volume-weighted average price (VWAP). Our experiments show that including these features consistently improves predictive performance across multiple ML architectures. We observe gains across several key metrics, including log-likelihood, mean squared error (MSE), $R^2$, conditional variance estimation, and directional accuracy.
\end{abstract}

\section{Introduction}
Intraday financial data offers a rich opportunity for building predictive models, yet the technical complexity of working with trading tick data presents unique challenges. Among the various ways to represent intraday activity, OHLC (Open, High, Low, Close) bar data remains one of the most widely used formats due to its balance between expressiveness and tractability. Compared to raw tick data, bar data provides a structured and compressed view of price movements over fixed intervals, making it easier to model dynamic market behavior. This trade-off—somewhat reduced expressiveness in exchange for vastly reduced complexity—makes bar data especially attractive in many practical trading systems and research applications. 

Despite its widespread use, most modeling efforts with bar data rely on a narrow set of features derived directly from the OHLC values and volume. In this paper, we explore a set of features introduced in a new intraday OHLC bar product by Bloomberg: the timestamps of the open, high, low, and close prices within each bar. To our knowledge, this type of timing information has not been systematically studied in the context of bar-level predictive modeling. Our goal is to evaluate whether incorporating this additional data can improve model performance on a common financial prediction task. 

We conduct a comprehensive empirical analysis using multiple machine learning models to compare standard OHLC-based features with extended versions that include the new timing data. Our experiments show that these features lead to consistent performance gains across different model architectures and evaluation metrics, indicating their potential as a valuable addition to the standard bar data toolkit. 

The rest of the paper is structured as follows: Section 2 reviews related work on bar data and intraday modeling. Section 3 introduces the new data features and outlines our methodology. Section 4 presents experimental results and analysis. Finally, Section 5 concludes and discusses directions for future research.

\section{Related Work}
Many papers have explored machine learning approaches using traditional OHLC bar data, including \cite{BANSAL2022247,10.1007/978-3-319-68612-7_56}. Other work has applied AI and forecasting techniques to tick data \cite{https://doi.org/10.1111/exsy.13537,10.1016/j.bdr.2023.100414} and order book information \cite{8713851}. 

However, to our knowledge, no systematic studies have yet examined the timing-enhanced OHLC data introduced here, as this dataset has only recently become available. In the next sections, we summarize the traditional OHLC structure and introduce the additional timing features used in this study.

\section{Methodology}
\subsection{OHLC Data}
\subsubsection{Traditional OHLC Bar Data}
Financial professionals have long relied on high-quality intraday data to inform their decision-making. Because working with the most granular form of this data—individual ticks—can be technically complex and resource-intensive, practitioners often prefer aggregated representations that summarize price behavior over fixed time intervals. One common format is the open-high-low-close (OHLC) bar, which records the opening, highest, lowest, and closing prices within a given time period, such as one minute. These datasets, often referred to as candlestick or OHLC data, typically contain minimal additional information beyond a security identifier, the timestamp of the interval, and the OHLC prices. In many cases, volume data is also included. Table ~\ref{tab:ohlcv_table} below presents a synthetic example of traditional OHLC data sampled at one-minute intervals.

\begin{table*}[ht]
\centering
\resizebox{\textwidth}{!}{%
    \begin{tabular}{lllrrrrr}
    \toprule
    \textbf{Ticker} & \textbf{Date} & \textbf{Time} & \textbf{Open} & \textbf{High} & \textbf{Low} & \textbf{Close} & \textbf{Volume (xRound Lots)} \\
    \midrule
    XYZ & 1/2/2023 & 9:30 AM EST & 100.00 & 100.12 & 99.94 & 99.96 & 1000 \\
    XYZ & 1/2/2023 & 9:31 AM EST & 99.95  & 100.08 & 99.87 & 100.08 & 500 \\
    XYZ & 1/2/2023 & 9:32 AM EST & 100.10 & 100.25 & 100.10 & 100.22 & 2000 \\
    \bottomrule
    \end{tabular}
}
\caption{Traditional OHLC bar data. The Volume is given in round lots of 100 shares.}
\label{tab:ohlcv_table}
\end{table*}

Such information is often presented in graphical form in what is commonly called a candlestick chart. We show the candlestick form of the traditional OHLC table in Figure ~\ref{fig:ohlcv_figure} below.

\begin{figure}[ht]
\centering
\includegraphics[width=0.8\linewidth]{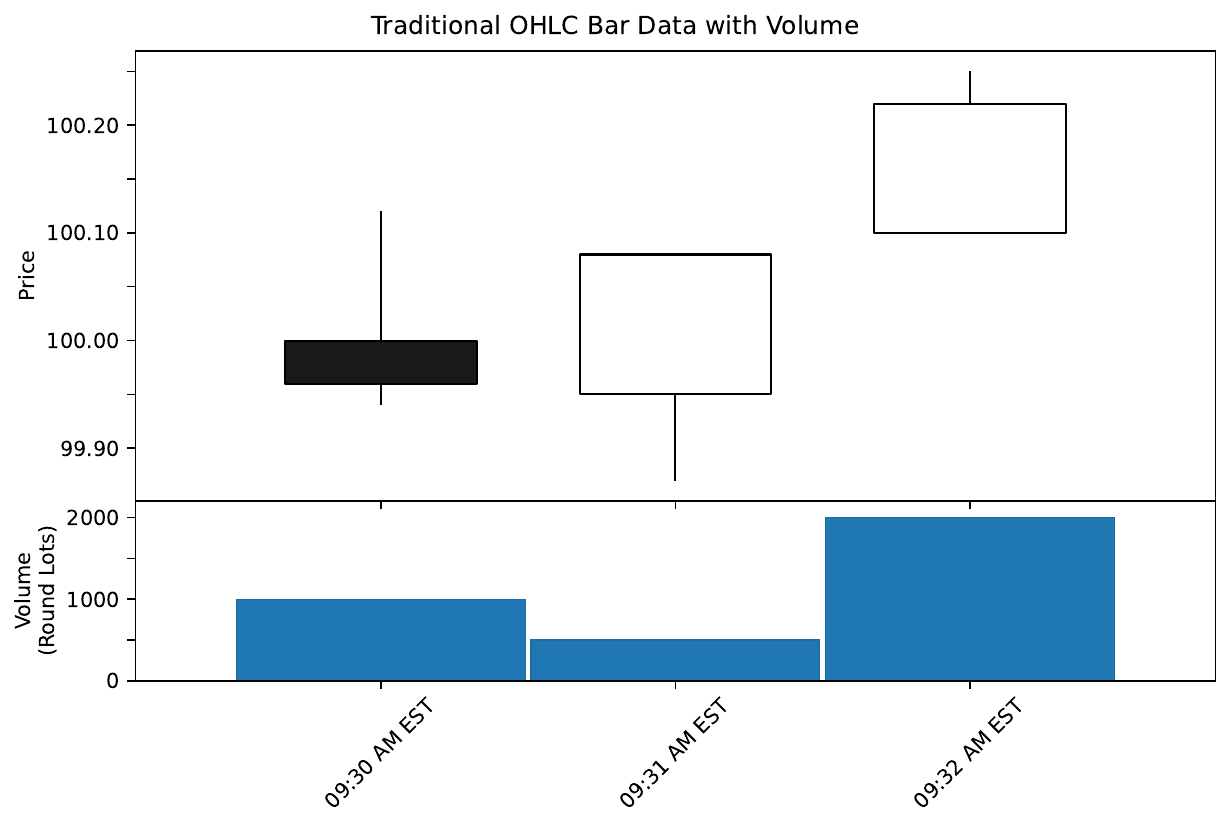}
\caption{Candlestick chart representation of traditional OHLC bar data with volume }
\label{fig:ohlcv_figure}
\end{figure}

\subsubsection{Enhancements to Traditional OHLC Data}
In 2024, Bloomberg launched a new service that allows enterprise clients to generate customized, high-quality OHLC bar data directly from tick data.\footnote{Original announcement available at {\url{https://www.bloomberg.com/company/press/bloomberg-launches-fully-customizable-intraday-quant-pricing-solution-for-investment-research/}}} This capability enables the inclusion of additional intra-bar information that enhances the utility of each bar. These enhancements fall into three main categories: 

\textbf{Timestamp Annotations}
Rather than reporting only the OHLC prices, the enhanced dataset also includes the exact timestamps at which these prices occurred within each interval. Specifically, it records the timestamp of the first trade for the \textit{Open} price, the first occurrence of the \textit{High} and \textit{Low} prices, and the last trade for the \textit{Close} price. 
These timestamp fields introduce a temporal dimension to each bar, supporting event-sequencing analysis and more nuanced modeling of intra-bar dynamics.

\textbf{VWAP Information}
The Bloomberg-generated bars include a volume-weighted average price (VWAP) calculated by Bloomberg for each interval. This execution-weighted metric serves as a useful complement to the standard OHLC values, providing insight into where the bulk of trading occurred during that period. The Bloomberg VWAP also functions as an industry-standard benchmark for evaluating trade execution quality. Importantly, the VWAP calculation can be customized by adjusting the trade condition codes included in the input, allowing for tailored definitions of execution quality and enabling the analyst to emphasize different aspects of price formation in a given modeling task.

\textbf{Market Microstructure Details}
Clients can configure the data to include or exclude trades based on specific condition codes, enabling cleaner filtering and more targeted analysis. Additionally, the dataset reports condition-code-specific volume and tick counts within each bar, offering granular insights into trade composition and market behavior. This is in addition to overall volume and tick count data—the latter representing a new field within the OHLC bar framework.

\subsection{Data Used}
\subsubsection{Universe}
For our analysis, we used Bloomberg’s intraday service to generate OHLC bar data for all members of the Russell 3000\textsuperscript{\textregistered} Index during the 2021 calendar year. We selected this index due to its broad coverage across industries, geographic exposure, and market capitalizations, making it a representative sample of U.S. equity markets. For each trading day in 2021, we restricted the dataset to that day’s official index constituents and included only trades executed during regular U.S. market hours and subject to the default set of trade condition codes.

To ensure data quality and relevance, we applied a set of filters to exclude bars with insufficient trading activity. Specifically, we removed bars in which the price or number of ticks fell below minimum thresholds, bars for which the target variable was unavailable, and bars with insufficient data or variation in their recent historical values. The full set of filtering rules is provided in Appendix \ref{appendix:data_filtering_rules}. 

These filters address two key concerns. First, when a stock exhibits limited trading activity during a given interval, its bar data can be dominated by a small number of trades, reducing the reliability of any statistical analysis. Second, when a stock's absolute price is very low, even modest price movements—such as those caused by crossing the bid-ask spread—can appear as disproportionately large returns relative to the previous trade.

\subsubsection{Training, Validation, and Test Sets}
We split the data temporally into three non-overlapping subsets. The training set includes data from the beginning of 2021 until August 9 2021. The validation set spans August 10--November 7 2021, and the test set covers November 8--December 31 2021. This split simulates a realistic forecasting setting and helps prevent data leakage across phases of model development.

\subsection{Choice of Target Variable}
Given a set of previous bars (the lookback period), our objective is to predict the relative change–a type of price return—in the VWAP value of the next, unseen bar. We focus on VWAP return for several reasons:
\begin{itemize}
    \item \textbf{Market relevance}: VWAP is widely followed by institutional investors, trading algorithms, and benchmark providers~\cite{HUMPHERYJENNER20112319}~\cite{BIALKOWSKI20081709}. As such, it is a meaningful and valuable quantity to predict.
    \item \textbf{Interval-wide robustness}: Unlike point-in-time prices such as the open or close, VWAP summarizes trading activity across the full interval, making it a more stable and representative measure of price.
    \item \textbf{Reduced discretization noise}: VWAP is a continuous, volume-weighted price and is therefore less affected by the rounding effects of penny-level price changes. In contrast, returns computed from last-trade prices (e.g., close-to-close) are often either exactly zero or artificially large due to the \$0.01 minimum tick increment.
    \item \textbf{Weaker arbitrage incentives}: Predicting the direction of VWAP changes does not lend itself to straightforward arbitrage. A trader who knows that the next bar's VWAP will be higher than the current one cannot directly profit from this knowledge unless they had already executed near the current VWAP. As a result, patterns in VWAP returns may persist even in broadly efficient markets.
\end{itemize}

\subsection{Lookback Window Size Choice}
For this analysis, we chose to fix a single lookback window size and use it consistently across all model trainings and experiments. In preliminary testing, we observed that longer lookback windows tended to yield more powerful models than shorter ones—even when accounting for the distributional shift that occurs with larger historical contexts. This finding supports the use of longer windows, as they appear to provide richer context for learning. 

However, longer lookback windows also reduce the number of usable time intervals, particularly at the start of each trading day. A model that requires $M$ previous one-minute bars cannot produce predictions during the first $M$ minutes of the day. Because the market open is often a period of heightened volatility and trading activity—arguably one of the most interesting windows to analyze—this limitation argues in favor of shorter lookback windows. 

To resolve this fundamental trade-off, we manually fixed the lookback window size to 20. This value was chosen as a compromise: long enough to capture meaningful context, but short enough to retain usability across the majority of the trading day. 

While it is also possible to build models that support variable or adaptive lookback windows, doing so would introduce substantial additional complexity into the modeling and evaluation process. We leave this line of investigation to future work.

\subsection{Feature Sets}
We trained separate models using three distinct sets of input features:

\textbf{Basic feature set}: Includes only the Open, High, Low, and Close prices, as well as past VWAP returns. VWAP return is included as a feature because it is also the target variable, and it would be unrealistic to model a return process without access to its historical values.

\textbf{No timing feature set}: Includes all available features except those derived from timing information (e.g., timestamps of high/low). This set serves as a rigorous baseline for evaluating the incremental predictive value of timing-based features.

\textbf{Full feature set}: Includes all available features, including those based on timing information. A full description of all features is provided in Appendix \ref{appendix:data_features}.

\subsection{Machine Learning Architectures}
\label{subsec:ml_arch_subsec}
We employed three standard machine learning architectures: multilayer perceptrons (MLPs), recurrent neural networks (RNNs), and Transformers with an attention mechanism. These architectures are widely used in time series and financial modeling, so we do not provide formal definitions here. Instead, we describe how input data was structured and processed within each architecture in the subsections below.

All architectures output the parameters of a Student’s $t$ distribution and are trained via maximum likelihood estimation. We selected the Student’s $t$ distribution because prior work has shown it to outperform the conditional normal distribution when modeling financial returns~\cite{10.1145/3677052.3698647}. We further ablate this choice in the Experiments section.

\subsubsection{Multilayer Perceptrons (MLPs)}
For the MLP architecture, we flatten the input sequence across both time and feature dimensions. Specifically, if the input consists of 20 one-minute bars, each with $D$ features, it is reshaped into a single vector of size $20 \times D$. This flattened vector is passed through a standard feedforward neural network, which outputs the parameters of a Student’s $t$ distribution.

\subsubsection{RNNs}
For the RNN architecture, the input sequence is preserved in its original form; there is no need for flattening. The sequence of one-minute bars in the lookback window is passed directly into a gated recurrent unit (GRU) network, which processes the input sequentially. The hidden state from the final time step is then passed into an MLP, which outputs the parameters of a Student’s $t$ distribution.

\subsubsection{Transformers}
For the Transformer architecture, we adopt a BERT-style approach in which a special classification token~\cite{devlin-etal-2019-bert} is prepended to the input sequence to serve as a summary representation of the entire sequence. Before entering the Transformer layers, the input sequence is passed through an initial MLP-based embedding network, analogous to the token embedding layers used in language models. This step projects each input vector—including the special token—into a common latent space. 

Each input is also augmented with a time embedding to encode its relative position within the lookback window, ensuring that the model is aware of temporal ordering. 

The full sequence, including the transformed special token and time-aware inputs, is then passed through a stack of Transformer encoder layers. After processing, the hidden state corresponding to the special token is passed into an MLP, which outputs the parameters of a Student’s $t$ distribution.

\subsection{Training Procedures}
\label{subsec:ml_training_proc_subsec}
Each of the three architectures was trained using each of the three feature sets. For every architecture–feature-set combination, hyperparameters were tuned to maximize log-likelihood on the validation set. Each hyperparameter configuration was evaluated across three random seeds, and final performance metrics were reported as the mean across these runs. In the sections below, we specify which hyperparameters were tuned for each architecture.

\subsubsection{MLP and RNN Tuning}
For the MLP and RNN architectures, we first tuned the dropout rate, followed by combinations of weight decay and learning rate. In preliminary experiments, other hyperparameters had minimal impact on validation loss and were therefore held constant. The specific values tested are provided in Table~\ref{tab:hyperparams_tuned_body} below. Additional hyperparameter settings are listed in Appendix \ref{appendix:manual_hyperparams}.

\begin{table}[ht]
\centering
\resizebox{\textwidth}{!}{%
\begin{tabular}{lccc}
\toprule
\textbf{Hyperparameter} & \textbf{MLP} & \textbf{RNN} & \textbf{Transformer} \\
Dropout rate & \texttt{\{0.1, 0.2, 0.3, 0.4, 0.5\}} & \texttt{\{0.1, 0.2, 0.3, 0.4, 0.5\}} & \texttt{\{0.1, 0.2, 0.3, 0.4, 0.5\}} \\
Weight decay & \texttt{\{1e-5, 1e-4, 1e-3, 1e-2\}} & \texttt{\{1e-5, 1e-4, 1e-3, 1e-2\}} & \texttt{\{1e-5, 1e-4, 1e-3, 1e-2\}} \\
Learning rate & \texttt{\{1e-5, 1e-4, 1e-3, 1e-2\}} & \texttt{\{1e-5, 1e-4, 1e-3, 1e-2\}} & \texttt{\{1e-6, 1e-5, 1e-4\}} \\
Initial embedding size & \texttt{N/A} & \texttt{N/A} & \texttt{\{32, 64, 128, 256\}} \\
Embedding MLP hidden size & \texttt{N/A} & \texttt{N/A} & \texttt{\{128, 256, 512\}} \\
Attention blocks & \texttt{N/A} & \texttt{N/A} & \texttt{\{4, 6, 8, 10, 12\}} \\
Attention heads & \texttt{N/A} & \texttt{N/A} & \texttt{\{2, 4, 6, 8\}} \\
\bottomrule
\end{tabular}
}
\caption{Search space for hyperparameters used in model tuning. \texttt{N/A} indicates the parameter is not applicable to that architecture.}
\label{tab:hyperparams_tuned_body}
\end{table}

\subsection{Target Distribution Across Splits}
\begin{figure}[H]
\centering
\includegraphics[width=0.8\linewidth]{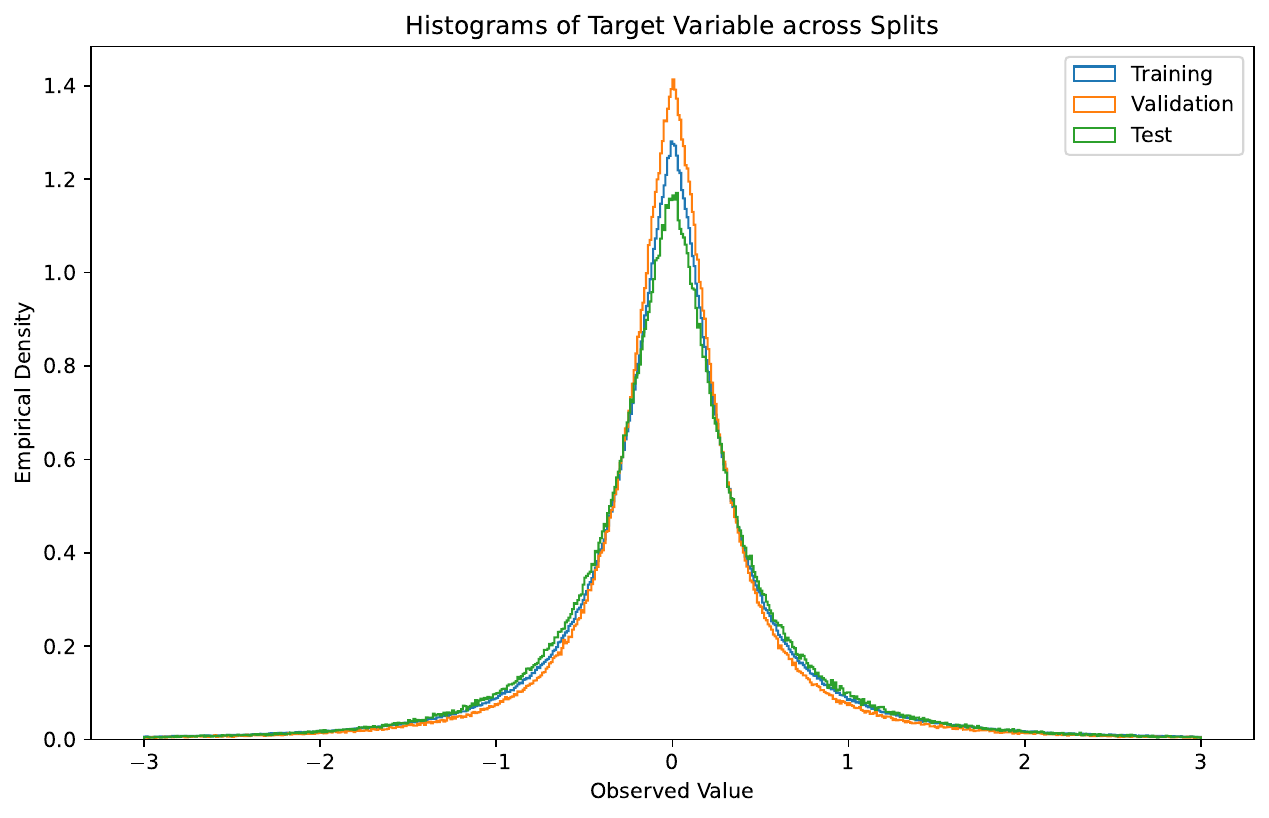}
\caption{\footnotesize Empirical density plots of the target variable across training, validation, and test splits. These represent the distribution of the unconditional target variable, in contrast to the conditional Student’s \(t\) distributions used for modeling predictive uncertainty. Note that the unconditional and conditional distributions are not expected to match.}
\label{fig:target_histograms}
\end{figure}

To characterize potential distributional shifts in the target variable (log of VWAP return), Table~\ref{tab:target_summary_stats} presents summary statistics for the training, validation, and test splits, while Figure~\ref{fig:target_histograms} shows histograms of the target values for each set.
\begin{table}[ht]
\centering
\resizebox{\textwidth}{!}{%
\begin{tabular}{l *{3}{S[scientific-notation=true,round-mode=places, round-precision=2]}}
\toprule
\textbf{Statistic} & \textbf{Training} & \textbf{Validation} & \textbf{Test} \\
\midrule
Mean & 0 & -1.75e-3 & -3.34e-3 \\
\textit{Median} & 2.08e-3 & 1.08e-3 & -2.40e-3 \\
\cmidrule(lr){1-4}
Standard deviation & 1.00e0 & 7.92e-1 & 8.53e-1 \\
\textit{Interquartile range} & 5.19e-1 & 4.57e-1 & 5.55e-1 \\
\cmidrule(lr){1-4}
Skewness & 3.74e-2 & -4.51e-2 & 1.55e-1 \\
\textit{Quartile skewness} & -4.90e-3 & -5.36e-3 & -4.10e-3 \\
\cmidrule(lr){1-4}
Excess kurtosis & 4.05e2 & 6.93e1 & 8.27e1 \\
\textit{Quantile excess kurtosis} & 7.15e0 & 6.56e0 & 5.11e0 \\
\bottomrule
\end{tabular}
}
\caption{\footnotesize Summary statistics of the target variable across training, validation, and test splits. The training set mean is effectively zero (within machine precision). Standard moment statistics are presented alongside robust alternatives (in italics). The robust statistics are more consistent across splits, highlighting the influence of extreme outliers on the standard measures. Quantile excess kurtosis is computed as the ratio of the 1\% trimmed range to the interquartile range, minus 3.449—the corresponding value for a Normal distribution.}
\label{tab:target_summary_stats}
\end{table}

The overall similarity of the density plots aligns with the strong agreement in the robust summary statistics. However, it is notable that the validation set appears significantly more peaked, and the test set less peaked, than the others—consistent with the variation in interquartile ranges. Although the test set has a smaller standard deviation than the training set, its larger interquartile range suggests greater intraday volatility. Among the three splits, the test set is also the only one with a negative median after normalization, indicating more bearish conditions. In the next section, we will examine how these distributional shifts influence the predictive performance of the machine learning models.

\section{Experiments and Results}
In this section, we present the performance of the three model architectures introduced in Section~\ref{subsec:ml_arch_subsec}—trained and tuned as described in Section~\ref{subsec:ml_training_proc_subsec}. Each architecture is evaluated across three different feature sets on a held-out validation set, using negative log-likelihood (NLL) as the primary metric. This comparison is used to identify the best-performing model configuration. Based on validation NLL, we select the top model for further evaluation. The remainder of this section focuses on that model’s performance on both the validation and test sets, where we assess its calibration, regression accuracy, directional prediction ability, and performance relative to simple baselines.

\subsection{Validation NLL}
\label{subsec:validation_nll}
Table~\ref{tab:nll_arch_results} reports the negative log-likelihood (NLL) on the validation set for each model architecture and feature set combination, using the best-performing hyperparameter configuration in each case.
\begin{table}[ht]
\centering
\small
\begin{tabular}{l *{3}{S[table-format=1.4,scientific-notation=fixed,round-mode=places, round-precision=4]}}
\toprule
\textbf{Feature Set} & \textbf{MLP} & \textbf{RNN} & \textbf{Transformer} \\
\midrule
Basic feature set & 0.2732 & \textit{0.2589} & 0.2667 \\
\cmidrule(lr){1-4}
No timing feature set & 0.2486 & 0.2460 & \textit{0.2450} \\
\cmidrule(lr){1-4}
Full feature set & 0.2480 & 0.2452 & \multicolumn{1}{S[table-format=1.4]}{\textbf{\textit{0.2442}}} \\
\bottomrule
\end{tabular}
\caption{Validation set negative log-likelihood (NLL) across model architectures and feature sets. The best model for each feature set is italicized; the best overall result is both italicized and bolded.}
\label{tab:nll_arch_results}
\end{table}

The best-performing model was the Transformer with all features, which achieved the lowest validation NLL using a dropout rate of 0.1, weight decay of $10^{-5}$, and learning rate of $10^{-4}$. This configuration included 10 attention blocks with 4 heads, an initial embedding size of 128, and an embedding MLP hidden size of 256. Detailed hyperparameter configurations for all architecture and feature set combinations are provided in Appendix \ref{appendix:tuned_hyperparams}. 

We focus on the Transformer architecture for the remainder of the evaluations. It consistently outperformed the other models on the richer feature sets, aligning with the main goal of this work: evaluating the added value of timing information. While its performance on the Basic feature set was relatively modest, that feature set is intended primarily as a baseline, not as a standalone input configuration.

\subsection{Test NLL, Baselines, and Ablation}
Table~\ref{tab:valid_vs_test_nll} reports the negative log-likelihood (NLL) on the validation and test sets for the best-performing Transformer models identified in Section 4.1. Since the target variable was standardized using the training set mean and variance, we also include the NLL of the test set under an unconditional standard normal distribution as a simple baseline for comparison.

\begin{table}[ht]
\centering
\small
\begin{tabular}{l *{2}{S[table-format=1.4,scientific-notation=fixed,round-mode=places, round-precision=4]}}
\toprule
\textbf{Feature Set} & \textbf{Validation Set} & \textbf{Test Set} \\
\midrule
Basic feature set & 0.2667 & 0.4311 \\
\cmidrule(lr){1-3}
No timing feature set & 0.2450 & 0.4088 \\
\cmidrule(lr){1-3}
Full feature set & \multicolumn{1}{S[table-format=1.4]}{\textbf{0.2442}} & \multicolumn{1}{S[table-format=1.4]}{\textbf{0.4075}} \\
\midrule
\midrule
Baseline $\mathcal{N}(0, 1)$ & 1.2329 & 1.2830 \\
\bottomrule
\end{tabular}
\caption{Negative log-likelihood (NLL) on validation and test sets for the best-performing Transformer models and a standard normal baseline. The lowest NLL for each dataset is shown in bold.}
\label{tab:valid_vs_test_nll}
\end{table}

Recall that our models output a Student’s $t$ distribution for each prediction. Table~\ref{tab:normal_student_nll_change} reports the change in negative log-likelihood (NLL) on the validation and test sets when replacing the full Student’s \(t\) distribution with a Gaussian that shares the same predicted mean and variance—effectively removing the degrees of freedom component and ignoring the heavy tails.

\begin{table}[ht]
\centering
\small
\sisetup{
  retain-explicit-plus,
  parse-numbers = true
}
\begin{tabular}{l *{2}{S[table-format=+1.4,scientific-notation=fixed,round-mode=places, round-precision=4]}}
\toprule
\textbf{Feature Set} & \textbf{Validation Set} & \textbf{Test Set} \\
\midrule
Basic feature set & +0.0271 & +0.0287 \\
\cmidrule(lr){1-3}
No timing feature set & +0.0255 & +0.0273 \\
\cmidrule(lr){1-3}
Full feature set & +0.0278 & +0.0297 \\
\bottomrule
\end{tabular}
\caption{Change in negative log-likelihood (NLL) on validation and test sets after removing the degrees of freedom from the Student’s \(t\) distribution.}
\label{tab:normal_student_nll_change}
\end{table}

Several observations emerge from these results. First, Table~\ref{tab:valid_vs_test_nll} shows a consistent improvement in NLL on both the validation and test sets as the feature set becomes more comprehensive. This suggests that the model is able to generalize the benefits of additional input information, particularly timing features, beyond the validation period. 

Second, the NLL values on the test set are noticeably higher than those on the validation set. Since the test set corresponds to a distinct time period, this discrepancy may reflect a shift in market conditions, with the test period exhibiting greater unpredictability. Further evidence of such a distribution shift appears in subsequent evaluations. 

Finally, Table~\ref{tab:normal_student_nll_change} shows a meaningful increase in NLL when the degrees of freedom component of the Student’s $t$ distribution is removed, effectively reducing it to a Gaussian with matched mean and variance. While this ablation does not include separate hyperparameter tuning for the Gaussian variant—so it is not a fully controlled comparison—it still provides evidence that the model is leveraging the heavy-tailed nature of the \textit{t}-distribution to improve predictive fit. 

Additional comparisons with alternative baselines—including a VWAP-to-close heuristic and a non-standardized zero-return model—are provided in Appendix \ref{appendix:compare_baselines}. These results further support the choice of the standardized $\mathcal{N}(0, 1)$ baseline, which outperforms the alternatives in MSE.

\subsection{Calibration Error}
One of the advantages of distributional prediction is that it provides an estimate of predictive uncertainty. Rather than outputting a single point estimate (e.g., the mean), a distributional model allows us to quantify the probability of different outcomes—for example, the likelihood that a predicted return is positive. 

To evaluate the model’s ability to capture uncertainty, we assess its calibration error. In the context of regression, perfect calibration is defined as:
\[
\mathbb{P}(Y < \mathcal{F}_X^{-1}(p)) = p, \quad \forall p \in [0, 1]
\]

where $\mathcal{F}_X$ is the predicted cumulative distribution function (CDF) for a given input $X$. In other words, the proportion of data points for which the model predicts a CDF value less than $p$ should be approximately $p$ itself. Following Kuleshov et al.~\cite{DBLP:journals/corr/abs-1807-00263}, we compute calibration error by comparing predicted quantiles to empirical coverage. The metric is defined as:
\begin{gather*}
\hat{p}_j = \left| \left\{ y_n \mid \mathcal{F}_{X_n}(y_n) < p_j,\; n = 1, \ldots, N \right\} \right| / N \\
\text{cal}(y_1, \ldots, y_N) = \sum_{j=1}^{M} \left( p_j - \hat{p}_j \right)^2
\end{gather*}

where ${\hat{p}}_j$ is the empirical fraction of observations below the $j$-th predicted quantile $p_j$, and $M$ is the number of quantiles evaluated. In this work, we use $M$=100 evenly spaced quantiles between 0 and 1.

Table~\ref{tab:calibration_results} reports the calibration error for each of the best-performing Transformer models from Section~\ref{subsec:validation_nll}, alongside both the unconditional $\mathcal{N}(0, 1)$ baseline and an ablated Gaussian variant that shares the mean and variance of the full feature set model’s Student’s $t$ distribution. To improve readability, all calibration error values have been multiplied by 100.

\begin{table}[ht]
\centering
\small
\begin{tabular}{l *{2}{S[table-format=1.4,scientific-notation=fixed,round-mode=places, round-precision=4]}}
\toprule
\textbf{Feature Set} & \textbf{Validation Set} & \textbf{Test Set} \\
\midrule
Basic feature set & 0.8438 & 1.3322 \\
\cmidrule(lr){1-3}
No timing feature set & 0.2236 & \multicolumn{1}{S[table-format=1.4]}{\textbf{0.0854}} \\
\cmidrule(lr){1-3}
Full feature set & \multicolumn{1}{S[table-format=1.4]}{\textbf{0.0330}} & 0.1148 \\
\midrule
\midrule
Baseline $\mathcal{N}(0, 1)$ & \multicolumn{1}{S[table-format=1.4e1]}{1.7377e2} & \multicolumn{1}{S[table-format=1.4e1]}{1.2468e2} \\
\cmidrule(lr){1-3}
Full feature set - Gaussian & 2.0310 & 1.5282  \\
\bottomrule
\end{tabular}
\caption{Calibration error ×100 on validation and test sets for the best-performing Transformer models, an unconditional baseline, and a Gaussian ablation. Large baseline values are shown in scientific notation for scale. The lowest calibration error for each dataset is shown in bold.}
\label{tab:calibration_results}
\end{table}

All Transformer models exhibit extremely low calibration error—several orders of magnitude smaller than the unconditional $\mathcal{N}(0, 1)$ baseline. While the full feature set model has slightly higher calibration error than the no timing feature set model on the test set, the difference is negligible in practice. In contrast, replacing the Student’s $t$ distribution with a Gaussian of equivalent mean and variance leads to a substantial increase in calibration error, underscoring the importance of modeling heavy tails. 

\begin{figure}[H]
\centering
\includegraphics[width=0.8\linewidth]{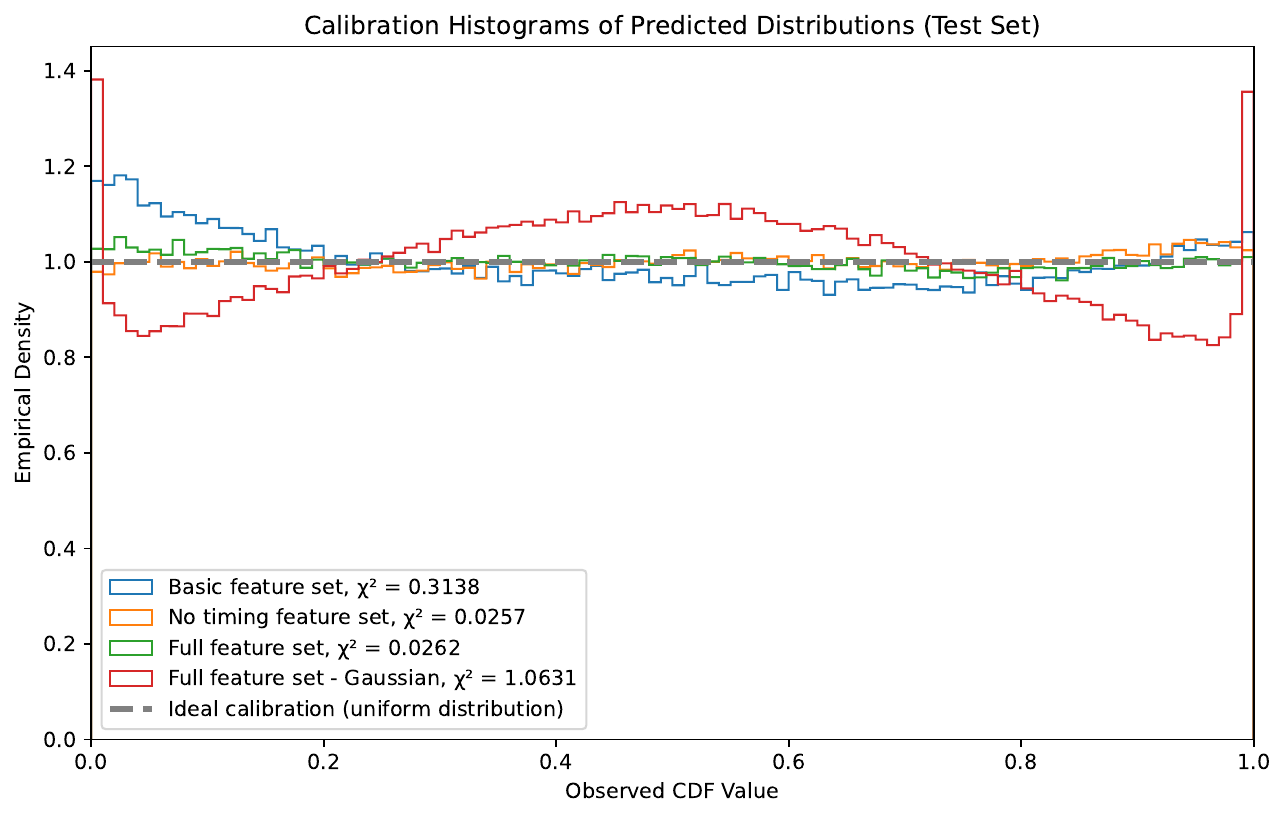}
\caption{\footnotesize Each curve shows the empirical distribution of CDF values evaluated at the observed outcomes for the best-performing Transformer model with a given feature set. The dashed line represents perfect calibration, while legend values indicate the Pearson chi-squared statistic, defined as the total sum of squared deviations from the uniform distribution over the 100 quantiles. Note that this statistic is used descriptively to summarize calibration error, not as a formal significance test.}
\label{fig:calibration_figure}
\end{figure}

Figure~\ref{fig:calibration_figure} illustrates these differences with histograms of predicted CDF values evaluated at the observed test outcomes. Under perfect calibration, these values should be uniformly distributed. Both the full feature set and no timing feature set models closely match this expectation, further confirming their reliability in capturing predictive uncertainty.

Both the full feature set and no timing feature set models exhibit good calibration, while substituting the Student’s \(t\) distribution with a Gaussian significantly degrades performance. The baseline $\mathcal{N}(0, 1)$ model is not shown in the figure, as its deviations from uniformity are so large that they distort the vertical scale.

\subsection{Mean Squared Error and \texorpdfstring{$R^2$}{R\textasciicircum2}}
Tables~\ref{tab:mse_results} and ~\ref{tab:r2_results} report the mean squared error (MSE) and coefficient of determination $R^2$ for the best-performing Transformer models identified in Section~\ref{subsec:validation_nll}, alongside the unconditional $\mathcal{N}(0, 1)$ baseline. In these evaluations, the predicted value for each input was taken to be the mean of the model’s output distribution.

\begin{table}[ht]
\centering
\small
\begin{tabular}{l *{2}{S[table-format=1.4,scientific-notation=fixed,round-mode=places, round-precision=4]}}
\toprule
\textbf{Feature Set} & \textbf{Validation Set} & \textbf{Test Set} \\
\midrule
Basic feature set & 0.4081 & 0.4717 \\
\cmidrule(lr){1-3}
No timing feature set & 0.3961 & 0.4602 \\
\cmidrule(lr){1-3}
Full feature set & \multicolumn{1}{S[table-format=1.4]}{\textbf{0.3942}} & \multicolumn{1}{S[table-format=1.4]}{\textbf{0.4555}} \\
\midrule
\midrule
Baseline $\mathcal{N}(0, 1)$ & 0.6279  & 0.7281 \\
\bottomrule
\end{tabular}
\caption{Mean squared error (MSE) on validation and test sets for the best-performing Transformer models and the baseline $\mathcal{N}(0, 1)$. The lowest MSE for each dataset is shown in bold.}
\label{tab:mse_results}
\end{table}

\begin{table}[ht]
\centering
\small
\begin{tabular}{l *{2}{S[table-format=1.4,parse-numbers=true]}}
\toprule
\textbf{Feature Set} & \textbf{Validation Set} & \textbf{Test Set} \\
\midrule
Basic feature set & 0.3500 & 0.3522 \\
\cmidrule(lr){1-3}
No timing feature set & 0.3691 & 0.3680 \\
\cmidrule(lr){1-3}
Full feature set & \multicolumn{1}{S[table-format=1.4]}{\textbf{0.3721}} & \multicolumn{1}{S[table-format=1.4]}{\textbf{0.3744}} \\
\midrule
\midrule
Baseline $\mathcal{N}(0, 1)$
  & \multicolumn{1}{S[scientific-notation=true, table-format=1.4e1]}{-4.8990e-6}
  & \multicolumn{1}{S[scientific-notation=true, table-format=1.4e1]}{-1.5345e-5} \\
\bottomrule
\end{tabular}
\caption{Coefficient of determination ($R^2$) on validation and test sets for the best-performing Transformer models and the baseline $\mathcal{N}(0, 1)$. The best $R^2$ in each dataset is bolded.}
\label{tab:r2_results}
\end{table}

All Transformer models outperform the baseline, which is effectively uninformative (with $\mathcal{N}(0, 1)$), and performance improves consistently as more features are added. Specifically, mean squared error (MSE) decreases and $R^2$ increases when moving from the basic feature set to the no timing feature set, and again with the addition of timing features.

Interestingly, while MSE and NLL degrade on the test set relative to the validation set, $R^2$ remains relatively stable. One possible explanation is that the models’ mean predictions remained accurate—capturing the direction and general scale of the targets—while test-time volatility increased, leading to higher MSE. The larger increase in NLL may then reflect a decline in the model’s ability to estimate predictive uncertainty, particularly the conditional variance. We explore this further in the next section. 

\begin{figure}[H]
\centering
\includegraphics[width=0.8\linewidth]{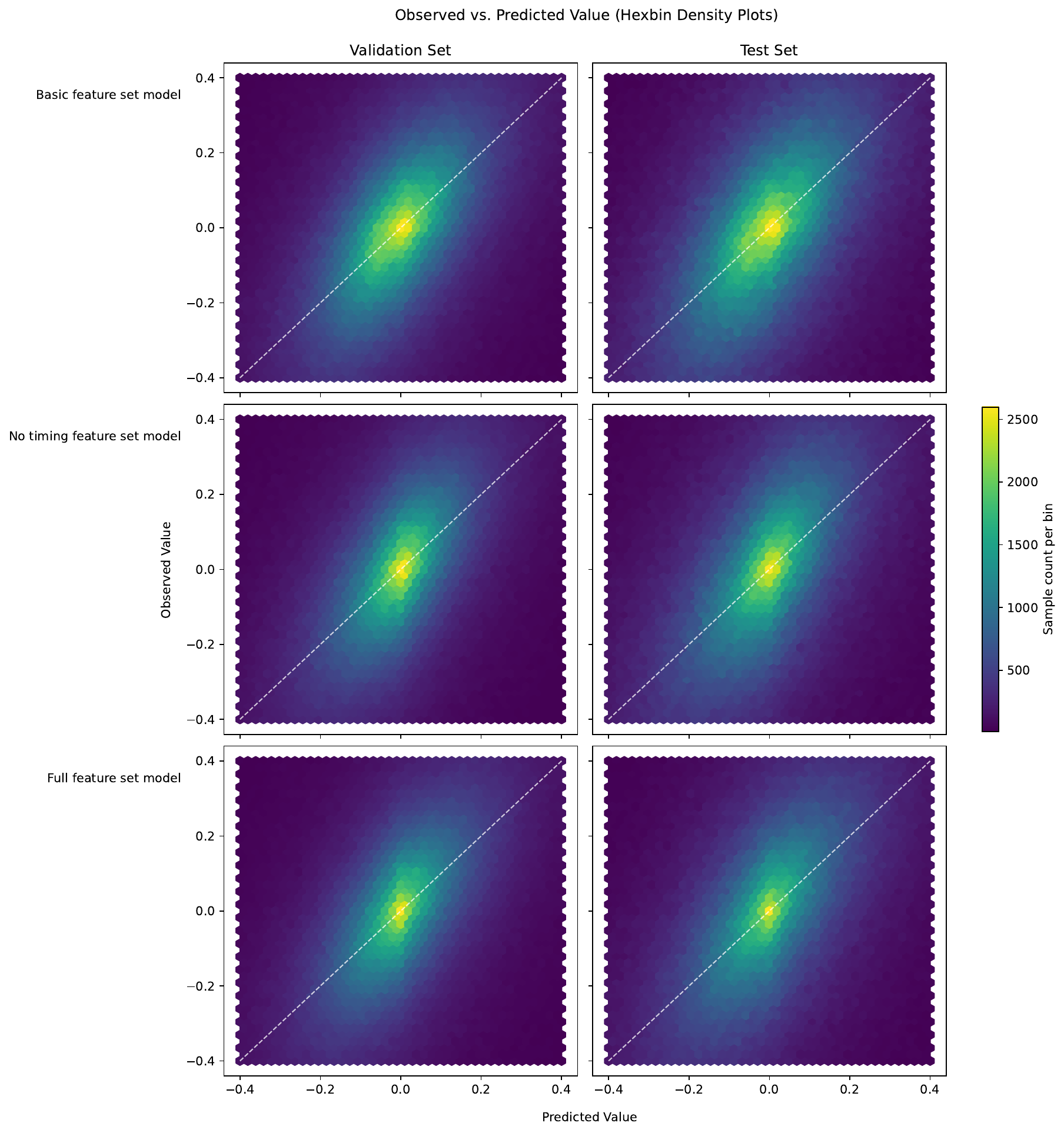}
\caption{\footnotesize Hexbin density plots of observed versus predicted values for the best-performing Transformer models. The white diagonal lines indicate perfect agreement between predictions and observations. The predicted values are less dispersed than the observations because only a portion of the total variance is due to variance in the conditional mean.}
\label{fig:mean_hexbin_density}
\end{figure}

To visually assess the performance of the models' mean predictions, Figure~\ref{fig:mean_hexbin_density} presents hexbin density plots of observed versus predicted values. The plots show that models with additional features produce tighter, more concentrated distributions around the diagonal, indicating more accurate predictions. In contrast, predictions on the test set appear more dispersed than those on the validation set, suggesting greater variability or a shift in the underlying data distribution.

\subsection{Conditional Variance}
In addition to negative log-likelihood, quantile calibration, and mean prediction, we also examine how well the model captures predictive variance structure. Although the squared error $(y-\mu)^2$ is a noisy estimate of the true conditional variance—especially under heavy-tailed distributions—it still provides useful qualitative insight. Results are reported in Table~\ref{tab:cond_var_results}.

\begin{table}[ht]
\centering
\small
\begin{tabular}{l *{2}{S[table-format=1.4]} S[table-format=2.2]}
\toprule
\textbf{Feature Set} & \textbf{Validation Set} & \textbf{Test Set} & \textbf{Increase (\%)} \\
\midrule
Basic feature set & 3.4825 & 4.4261 & 27.09\\
\cmidrule(lr){1-4}
No timing feature set & 3.4958 & 4.4752 & 28.02\\
\cmidrule(lr){1-4}
Full feature set & \multicolumn{1}{S[table-format=1.4]}{\textbf{3.3261}} & \multicolumn{1}{S[table-format=1.4]}{\textbf{4.2880}} & 28.92\\
\midrule
\midrule
Baseline $\mathcal{N}(0, 1)$ & 5.3145  & 6.7052 & 26.17\\
\bottomrule
\end{tabular}
\caption{Root mean squared error (RMSE) between predicted variance and realized squared error for the best-performing Transformer models and the baseline $\mathcal{N}(0, 1)$. The lowest RMSE for each dataset is shown in bold.}
\label{tab:cond_var_results}
\end{table}

The full feature set model clearly performs best in predicting conditional variance, while the no timing feature set model surprisingly underperforms the basic feature set model. This is unexpected, given that the additional non-timing features consistently improved performance on other metrics. 

The results also provide further evidence of a distribution shift in the test set. All models—including the baseline unconditional Normal—show a similar relative increase in RMSE from validation to test. This supports the hypothesis that the performance gap is primarily due to a secular increase in the volatility of volatility during the test period, rather than an issue with the training process such as hyperparameter overfitting. To complement the RMSE-based evaluation, we visualize both the distributions and the relationship between predicted and realized variances in Figures~\ref{fig:var_marginal_shifts} and ~\ref{fig:var_hexbin_shifts} below.

\begin{figure}[H]
\centering
\includegraphics[width=0.7\linewidth]{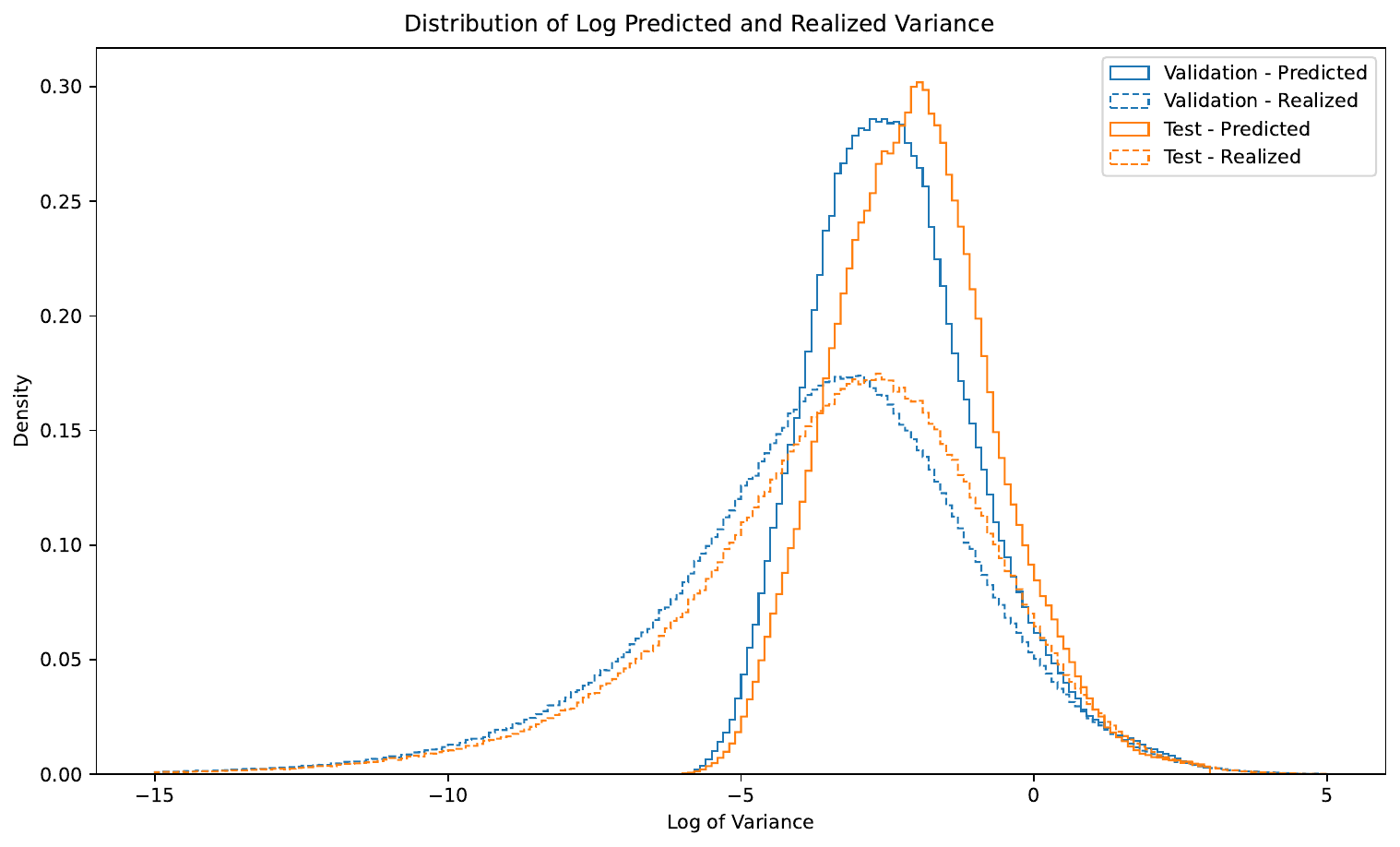}
\caption{\footnotesize Density plots of log predicted variance (solid) and log realized variance (dashed) for the validation and test sets for the full feature set Transformer model. A visible rightward shift in the test set curves reflects increased uncertainty and volatility in test conditions.}
\label{fig:var_marginal_shifts}
\end{figure}

\begin{figure}[H]
\centering
\includegraphics[width=0.7\linewidth]{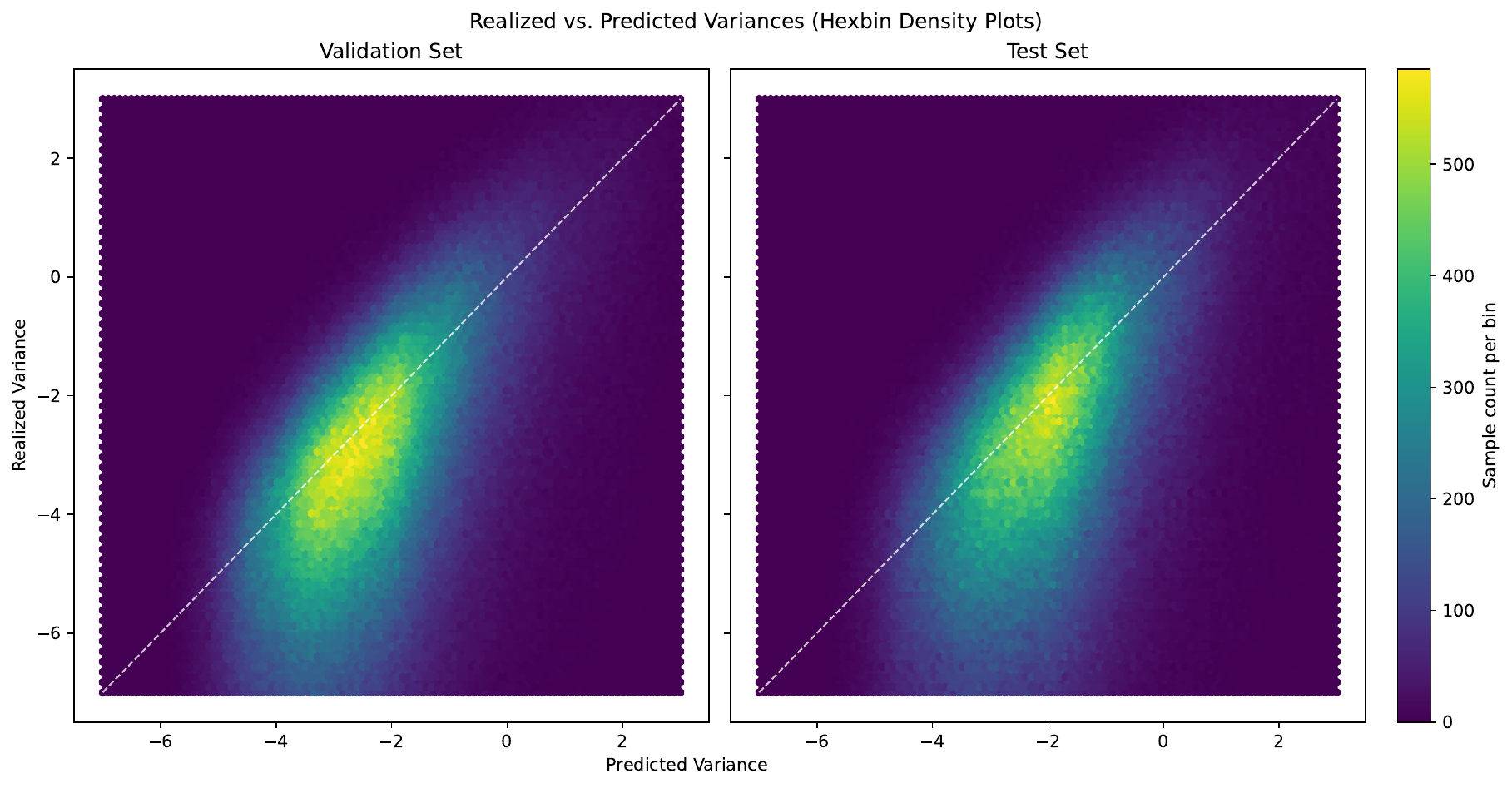}
\caption{\footnotesize 2D hexbin density plots comparing log predicted variance to log realized variance on the validation set (left) and test set (right) for the full feature set Transformer model. The white dashed line indicates perfect agreement. The test set shows a wider and slightly right-shifted distribution, consistent with increased predictive uncertainty and conditional variance mismatch.}
\label{fig:var_hexbin_shifts}
\end{figure}

\subsection{Directional Accuracy}
We evaluate the directional accuracy of the best-performing Transformer models by measuring how often the sign of the model’s predicted mean matches the sign of the observed value. Table~\ref{tab:direct_acc_overall} reports overall directional accuracy on the validation and test sets for each model. Table~\ref{tab:direct_acc_decile} shows the directional accuracy for each decile of the predicted absolute value for the full feature set model. Larger deciles represent model means that have relatively large absolute values, so they should represent predictions that are more directionally confident, and hence have higher accuracies. A similar per-decile analysis can also be performed from the perspective of observed (rather than predicted) value deciles. While detailed per-decile results for all models are provided in Appendix \ref{appendix:directional_accuracy_full}, we present only the full feature set model here, as all models exhibit similar patterns. Baseline values are not presented because the baseline $\mathcal{N}(0, 1)$ model does not predict any directions.

\begin{table}[ht]
\centering
\small
\begin{tabular}{l *{2}{S[table-format=2.3]}}
\toprule
\textbf{Feature Set} & \textbf{Validation Set Accuracy (\%)} & \textbf{Test Set Accuracy (\%)} \\
\midrule
Basic feature set & 71.750 & 71.889 \\
\cmidrule(lr){1-3}
No timing feature set & 72.106 & 72.288 \\
\cmidrule(lr){1-3}
Full feature set & \multicolumn{1}{S[table-format=2.3]}{\textbf{72.147}} & \multicolumn{1}{S[table-format=2.3]}{\textbf{72.294}} \\
\bottomrule
\end{tabular}
\caption{Directional prediction accuracy (\%) on validation and test sets for the best-performing Transformer models. An extra digit is shown to highlight small differences. The highest accuracy in each dataset is shown in bold.}
\label{tab:direct_acc_overall}
\end{table}

\begin{table}[ht]
\centering
\small
\begin{tabular}{l *{2}{S[table-format=2.3]}}
\toprule
\textbf{Decile} & \textbf{Validation Set Accuracy (\%)} & \textbf{Test Set Accuracy (\%)} \\
\midrule
1\textsuperscript{st} & 52.82 & 52.85 \\
2\textsuperscript{nd} & 58.20 & 58.58 \\
3\textsuperscript{rd} & 63.14 & 63.61 \\
4\textsuperscript{th} & 67.69 & 67.85 \\
5\textsuperscript{th} & 71.75 & 71.76 \\
6\textsuperscript{th} & 75.42 & 75.38 \\
7\textsuperscript{th} & 78.50 & 78.89 \\
8\textsuperscript{th} & 81.70 & 81.41 \\
9\textsuperscript{th} & 84.53 & 84.40 \\
10\textsuperscript{th} & 87.72 & 88.21 \\
\midrule
\midrule
Overall & 72.15 & 72.29 \\
\bottomrule
\end{tabular}
\caption{Directional prediction accuracy (\%) for the full feature set Transformer model, reported by decile of the absolute predicted means.}
\label{tab:direct_acc_decile}
\end{table}

The directional accuracies reported in Table~\ref{tab:direct_acc_overall} increase with larger feature sets, indicating that the additional inputs improve the model's ability to predict the correct direction of change. To better quantify the impact, it is helpful to consider the ratio of correct to incorrect predictions. On the validation set, this ratio improves from 2.54:1 with the basic feature set to 2.59:1 with the full feature set—an increase of approximately 2\% before considering the decile effects. 

Table~\ref{tab:direct_acc_decile} further breaks down directional accuracy by decile of the predicted absolute value for the full feature set model. The results show a clear trend: the larger the predicted magnitude, the higher the likelihood that the predicted direction is correct. This has practical implications, as more confident predictions (i.e., those with higher absolute values) support taking larger positions in downstream applications. Notably, even in the lowest decile, directional accuracy remains above 50\%, suggesting that the model retains some predictive power even when its confidence is lowest.

\section{Discussion and Conclusion}
Our results show that timing information improves model performance on the primary evaluation metric—log-likelihood—across all machine learning architectures tested. Additionally, the best-performing architecture (Transformers) demonstrated consistent gains across key metrics, including mean squared error, $R^2$, conditional variance estimation, and directional accuracy. These findings suggest that timestamps associated with individual price points carry meaningful signal that can enhance intraday prediction tasks based on OHLC bar data.

Future work includes expanding the analysis to a broader range of securities and time periods, exploring alternative model architectures, and incorporating mechanisms to capture cross-asset correlations. 

Overall, our study demonstrates that the timing of the high and low prices in 1-minute bars has a significant impact on the predictability of VWAP, even when conventional features are already included. This analysis was enabled by the recent introduction of Bloomberg’s new intraday data product, which includes these timing fields—information that was previously unavailable without resorting to raw tick data. We hope these findings highlight the potential of this newly accessible intraday information to enhance predictive models, while preserving the tractability advantages of OHLC bar data. 

\bibliographystyle{plain}
\bibliography{references}

\appendix
\section{Data Filtering Rules}
\label{appendix:data_filtering_rules}
Besides selecting point-in-time index members of the Russell 3000\textsuperscript{\textregistered} Index during U.S. trading hours, all of the data was additionally filtered the same way for the training, validation, and test sets. The filtering rules are given in Table~\ref{tab:filtering_rules} below. These rules were developed during exploratory data analysis with the goal of keeping as much data as possible while removing data points that were too unreliable to say anything about.

\begin{table}[ht]
\centering
\resizebox{\textwidth}{!}{
    \begin{tabular}{l l}
    \toprule
    \textbf{Feature(s)} & \textbf{Condition}\\
    \midrule
    Minimum low price during the lookback window & $\geq$ \$4$^1$ \\
    \cmidrule(lr){1-2}
    Minimum number of ticks per bar during the lookback window & $\geq$ 30$^2$ \\
    \cmidrule(lr){1-2}
    High and low prices during the lookback window & $\max(high)$ $\ne$ $\min(low)$ \\
    \cmidrule(lr){1-2}
    Open, High, Low, and Close prices during the lookback window & $\ne$ NaN \\
    \cmidrule(lr){1-2}
    Observed target variable (in our case, log of VWAP return) & $\ne$ NaN \\
    \bottomrule
    \end{tabular}
}
\caption{Filtering rules for data.\\
$^1$ This is the minimum bid price for a security to become listed on NASDAQ. Low prices also amplify the relative impact of small price changes, such that even crossing the spread can appear to be a large return. \\
$^2$ Bar-level features represent aggregate statistics computed from individual ticks. The threshold of 30 ticks was chosen manually to balance statistical robustness with data availability.}
\label{tab:filtering_rules}
\end{table}

After filtering, the validation set contained approximately 1.3 million entries spanning 1,021 unique stocks over 63 days. This implies that roughly two-thirds of all stocks were excluded by the filtering rules during the validation period. If every included stock had a valid observation for every possible minute bar over those 63 days (accounting for a 20-bar lookback window), the validation set would have been about 18 times larger. This highlights the substantial impact of the filtering rules, even after accounting for typically inactive stocks.

While this is not ideal—since a robust system should ideally generalize to both inactive and active stocks—any approach that relies on trading activity to extract meaningful insights will, in practice, require special handling for cases with minimal or no activity.

\section{Feature Set Description}
\label{appendix:data_features}
The full set of data features used for modeling is given below in Table~\ref{tab:full_feature_dict}.
\begin{table}[H]
\centering
\resizebox{\textwidth}{!}{
    \begin{tabular}{l l p{10cm}}
    \toprule
    \textbf{Feature Group (Normalization)} & \textbf{Feature} & \textbf{Definition}\\
    \midrule
    \multicolumn{3}{l}{\textbf{Basic Prices} (min–max$^1$)}\\ & Open, High, Low, and Close & The first, highest, lowest, and last prices in one bar, respectively.\\
    \addlinespace[2pt]
    \multicolumn{3}{l}{\textbf{Log Returns} (Mean/Std. Dev. over the training set)} \\
    &VWAP Log Return & Log of the ratio of VWAP value in one bar to the one before.\\
    &Bar Log Return & Log of the ratio of close to open values within one bar.\\
    &High/Close Log Return & Log of the ratio of close to high values within one bar.\\
    &High/Low Log Return & Log of the ratio of high to low values within one bar.\\
    &Close/VWAP Log Return & Log of the ratio of close to VWAP values within one bar.\\
    \addlinespace[2pt]
    \multicolumn{3}{l}{\textbf{Volume Measures} (Median/IQR over the training set)} \\
    & Volume / Log Volume & The number of shares traded in one bar and its natural logarithm.\\
    & Dollar Volume / Log Dollar Volume & The VWAP times the Volume in one bar and its natural logarithm.\\
    & Tick Count & The number of individual ticks in one bar.\\
    \addlinespace[2pt]
    \multicolumn{3}{l}{\textbf{Bar Scale Measures} (Not normalized)} \\
    & Scaled Bar Height & The difference between the High and Low prices after min–max scaling the Basic Prices.\\
    & Scaled Close vs. Open & The difference between Close and Open prices after min–max scaling the Basic Prices.\\
    & Close Fraction & The Close price on a linear ($\in [0,1]$) scale between the Low and High within one bar.\\
    & Open Fraction & The Open price on a linear ($\in [0,1]$) scale between the Low and High within one bar.\\
    \addlinespace[2pt]
    \multicolumn{3}{l}{\textbf{Time of Day} (Not normalized)} \\
    & Minute Index & A scaled time of day as an integer; $0$ corresponds to the first trading minute of the day.\\
    \addlinespace[2pt]
    \multicolumn{3}{l}{\textbf{Volume In Recent Past} (Mean/Std. Dev. over the training set)} \\
    & Mean Prior Volume & Mean of the log dollar volume for this stock in this same minute over the last 5 days.$^2$\\
    & Std. Dev. Prior Volume & Standard deviation of the log dollar volume for this stock in this same minute over the last 5 days.\\
    & Median Prior Volume & Median of the log dollar volume for this stock in this same minute over the last 5 days.\\
    & Pct25 Prior Volume & 25\textsuperscript{th} percentile of the log dollar volume for this stock in this same minute over the last 5 days.\\
    & Pct75 Prior Volume & 75\textsuperscript{th} percentile of the log dollar volume for this stock in this same minute over the last 5 days.\\
    \multicolumn{3}{l}{\textbf{Relative Activity} (Not normalized)} \\
    & Prior Volume Standard Z Score & The volume in this bar, standardized by the Mean Prior Volume and Std. Dev. Prior Volume as defined above.\\
    & Prior Volume Robust Z Score & The volume in this bar, standardized by the Median Prior Volume, Pct25, and Pct75 Prior Volumes as defined above.\\
    \addlinespace[2pt]
    \multicolumn{3}{l}{\textbf{Basic Timing Features} (Scaled to [0,1]$^3$)}\\
    & High Time (normalized) & The first timestamp of the High price in the bar.\\
    & Low Time (normalized) & Similar to the High Time (normalized) above, but for the Low price.\\
    \addlinespace[2pt]
    \multicolumn{3}{l}{\textbf{Derived Timing Features} (Not normalized)} \\
    & Time Difference & The difference between High Time (normalized) and Low Time (normalized).\\
    & Timing Surprise & An indicator variable that is \texttt{True} if the order of the high and low times does not match the direction of the bar$^4$.\\
    \bottomrule
    \end{tabular}
}
\caption{\footnotesize Summary of all engineered features used in the model, grouped by type and normalized appropriately.\\
$^1$ These features are linearly rescaled within each data instance such that the lowest low in the lookback window maps to $-1$ and the highest high in the lookback window maps to $+1$. This approach is analogous to how plotting software scales data to fill the vertical axis, regardless of absolute price levels. Normalization is applied independently for each data instance.\\
$^2$ This is the length of a standard work week. This is meant to capture recent, as opposed to long-term activity.\\
$^3$ Timestamps are scaled such that $0$ is the time at the start of the bar and $1$ is the time at the end.\\
$^4$ A bar that goes up (the close is higher than the open) is much more likely to have the high time occur after the low time. We have investigated timing surprise informally in the past. Here we simply use it as a feature.\\
}
\label{tab:full_feature_dict}
\end{table}

\section{Manually Set Hyperparameters}
\label{appendix:manual_hyperparams}
In addition to the tuned hyperparameters, we made other architectural choices for each ML architecture based on our own previous work and on exploratory model analysis. The notable ones are highlighted in Table~\ref{tab:manual_hyperparams} below.

\begin{table}[ht]
\centering
\resizebox{\textwidth}{!}{
    \begin{tabular}{l l l l}
    \toprule
    \textbf{Hyperparameter} & \textbf{MLP} & \textbf{RNN} & \textbf{Transformer}\\
    \midrule
    Number of layers & 2 & 2 for the RNN and output MLP & 2 for the embedding and output MLPs \\
    \cmidrule(lr){1-4}
    Hidden sizes & 256 & 256 for the RNN and the output MLP & 256 for both the Transformer blocks and the output MLP \\
    \cmidrule(lr){1-4}
    Residual blocks & Yes & Yes for the RNN and the output MLP & Yes for the embedding and output MLPs \\
    \cmidrule(lr){1-4}
    Layer norms & Yes & Yes for the output MLP & Yes for the embedding and output MLPs \\
    \cmidrule(lr){1-4}
    Post-norm & Yes & Yes for the output MLP & Yes for the embedding and output MLPs \\
    \cmidrule(lr){1-4}
    Activation function & ReLU & ReLU for the output MLP, SiLU in the residual block & ReLU \\
    \cmidrule(lr){1-4}
    Training batch size & 1024 & 1024 & 1024 \\
    \cmidrule(lr){1-4}
    Epochs & 50 & 50 & 50 \\
    \cmidrule(lr){1-4}
    Batches per epoch & 1000 & 1000 & 1000 \\
    \bottomrule
    \end{tabular}
}
\caption{Manually set hyperparameter choices per architecture.}
\label{tab:manual_hyperparams}
\end{table}

\section{Hyperparameter Tuning Results}
\label{appendix:tuned_hyperparams}
Tables~\ref{tab:hyperresults_mlp}, ~\ref{tab:hyperresults_rnn}, and ~\ref{tab:hyperresults_transformer} below give the tuned hyperparameters for all combinations of model architectures and feature sets.

\begin{table}[H]
\centering
\resizebox{\textwidth}{!}{
    \begin{tabular}{l l l l}
    \toprule
    \textbf{Hyperparameter} & \textbf{Basic Feature Set} & \textbf{No Timing Feature Set} & \textbf{Full Feature Set} \\
    \midrule
    Dropout rate & 0.1 & 0.3 & 0.3 \\
    \cmidrule(lr){1-4}
    Weight decay & $10^{-2}$ & $10^{-2}$ & $10^{-2}$ \\
    \cmidrule(lr){1-4}
    Learning rate & $10^{-4}$ & $10^{-4}$ & $10^{-4}$ \\
    \bottomrule
    \end{tabular}
}
\caption{Selected hyperparameter values for the best MLP model per feature set.}
\label{tab:hyperresults_mlp}
\end{table}

\begin{table}[H]
\centering
\resizebox{\textwidth}{!}{
    \begin{tabular}{l l l l}
    \toprule
    \textbf{Hyperparameter} & \textbf{Basic Feature Set} & \textbf{No Timing Feature Set} & \textbf{Full Feature Set} \\
    \midrule
    Dropout rate & 0.1 & 0.1 & 0.2 \\
    \cmidrule(lr){1-4}
    Weight decay & $10^{-2}$ & $10^{-4}$ & $10^{-3}$ \\
    \cmidrule(lr){1-4}
    Learning rate & $10^{-4}$ & $10^{-4}$ & $10^{-4}$ \\
    \bottomrule
    \end{tabular}
}
\caption{Selected hyperparameter values for the best RNN model per feature set.}
\label{tab:hyperresults_rnn}
\end{table}

\begin{table}[H]
\centering
\resizebox{\textwidth}{!}{
    \begin{tabular}{l l l l}
    \toprule
    \textbf{Hyperparameter} & \textbf{Basic Feature Set} & \textbf{No Timing Feature Set} & \textbf{Full Feature Set} \\
    \midrule
    Dropout rate & 0.1 & 0.1 & 0.1 \\
    \cmidrule(lr){1-4}
    Weight decay & $10^{-2}$ & $10^{-4}$ & $10^{-5}$ \\
    \cmidrule(lr){1-4}
    Learning rate & $10^{-4}$ & $10^{-4}$ & $10^{-4}$ \\
    \cmidrule(lr){1-4}
    Initial embedding size & 32 & 32 & 128 \\
    \cmidrule(lr){1-4}
    Embedding MLP hidden size & 256 & 512 & 256 \\
    \cmidrule(lr){1-4}
    Attention blocks & 12 & 6 & 10 \\
    \cmidrule(lr){1-4}
    Attention heads & 8 & 4 & 4 \\
    \bottomrule
    \end{tabular}
}
\caption{Selected hyperparameter values for the best Transformer model per feature set.}
\label{tab:hyperresults_transformer}
\end{table}

\section{Comparative Baseline Evaluation}
\label{appendix:compare_baselines}
In this section, we evaluate several alternative baselines to justify our use of the $\mathcal{N}(0, 1)$ baseline throughout the earlier experiments. Recall that the target variable (log VWAP return) was standardized using the training set mean and variance, so a $\mathcal{N}(0, 1)$ baseline corresponds to a Gaussian distribution with those same parameters.

We consider two intuitive alternatives. The first sets the mean to exactly zero, rather than using the empirical mean of the training data. The second is based on the assumption that the price process behaves as a martingale—suggesting that the best predictor of the next bar’s price is the closing price of the previous bar. Under this assumption, the bar-to-bar VWAP-to-VWAP return could be approximated by the previous bar’s VWAP-to-close return.

Table~\ref{tab:baseline_comparison} compares the mean squared error (MSE) of these baselines alongside the full feature set Transformer model. Note that the MSE values in this table may differ slightly from those in Table~\ref{tab:mse_results}, as this comparison is restricted to instances where the VWAP-to-close return was computable, which represents a subset of the full dataset.

\begin{table}[ht]
\centering
\small
\begin{tabular}{l *{2}{S[table-format=1.5]}}
\toprule
\textbf{Baseline} & \textbf{Validation Set} & \textbf{Test Set} \\
\midrule
Zero return (normalized) & \multicolumn{1}{S[table-format=1.5]}{\textbf{0.62892}} & \multicolumn{1}{S[table-format=1.5]}{\textbf{0.72669}} \\
\cmidrule(lr){1-3}
Zero return (raw scale) & 0.62893 & 0.72670 \\
\cmidrule(lr){1-3}
Previous VWAP-to-close return & 0.88756 & 1.04104 \\
\midrule
\midrule
Full feature set Transformer model & 0.39551 & 0.45534 \\
\bottomrule
\end{tabular}
\caption{Mean squared error (MSE) on validation and test sets for various baselines. An extra digit is shown to highlight small differences. The lowest MSE per dataset is bolded. The Transformer model (full feature set) is included for reference.}
\label{tab:baseline_comparison}
\end{table}

These results confirm that our choice of baseline was reasonable: the normalized zero-return baseline slightly outperforms the non-normalized version, though the difference in MSE is quite small—on the order of $10^{-5}$. Perhaps more notably, the previous bar’s close performs poorly as a predictor of the next bar’s VWAP. This likely reflects the effects of bid-ask bounce and the inherent noise in price movements at the one-minute timescale.
\section{Directional Accuracy by Decile}
\label{appendix:directional_accuracy_full}
We showed the directional accuracy by decile of the full feature set Transformer model in Table~\ref{tab:direct_acc_decile}. Here we present the corresponding results for the basic feature set (Table~\ref{tab:direct_acc_decile_basic}) and no timing feature set (Table~\ref{tab:direct_acc_decile_notime}) models. The results show the same pattern of increasing accuracy per decile while maintaining a directional accuracy over 50\% for the lowest decile. As before, the deciles are computed from the absolute values of the predicted means.

\begin{table}[ht]
\centering
\small
\begin{tabular}{l *{2}{S[table-format=2.3]}}
\toprule
\textbf{Decile} & \textbf{Validation Set Accuracy (\%)} & \textbf{Test Set Accuracy (\%)} \\
\midrule
1\textsuperscript{st} & 52.64 & 52.62 \\
2\textsuperscript{nd} & 57.64 & 58.15 \\
3\textsuperscript{rd} & 63.11 & 63.28 \\
4\textsuperscript{th} & 67.28 & 68.11 \\
5\textsuperscript{th} & 71.64 & 71.54 \\
6\textsuperscript{th} & 74.80 & 75.05 \\
7\textsuperscript{th} & 78.25 & 78.00 \\
8\textsuperscript{th} & 80.89 & 80.90 \\
9\textsuperscript{th} & 84.16 & 83.76 \\
10\textsuperscript{th} & 87.09 & 87.47 \\
\midrule
\midrule
Overall & 71.75 & 71.89 \\
\bottomrule
\end{tabular}
\caption{Directional prediction accuracy (\%) for the basic feature set Transformer model, reported by decile of the absolute predicted means.}
\label{tab:direct_acc_decile_basic}
\end{table}

\begin{table}[ht]
\centering
\small
\begin{tabular}{l *{2}{S[table-format=2.3]}}
\toprule
\textbf{Decile} & \textbf{Validation Set Accuracy (\%)} & \textbf{Test Set Accuracy (\%)} \\
\midrule
1\textsuperscript{st} & 52.82 & 52.91 \\
2\textsuperscript{nd} & 58.36 & 58.77 \\
3\textsuperscript{rd} & 63.34 & 64.01 \\
4\textsuperscript{th} & 67.81 & 68.04 \\
5\textsuperscript{th} & 71.95 & 71.91 \\
6\textsuperscript{th} & 75.46 & 75.60 \\
7\textsuperscript{th} & 78.50 & 78.58 \\
8\textsuperscript{th} & 81.60 & 81.43 \\
9\textsuperscript{th} & 84.25 & 84.15 \\
10\textsuperscript{th} & 86.98 & 87.49 \\
\midrule
\midrule
Overall & 72.11 & 72.29 \\
\bottomrule
\end{tabular}
\caption{Directional prediction accuracy (\%) for the no timing feature set Transformer model, reported by decile of the absolute predicted means.}
\label{tab:direct_acc_decile_notime}
\end{table}

\end{document}